\documentclass{elsart}

\usepackage{epsfig,subfigure,latexsym,amsmath}
\usepackage{amssymb}
\usepackage{epsfig}
\usepackage{rotating}

\begin{document}

\input epsf

\begin{frontmatter}

\title{Predictions of $\alpha$-decay half-lives based on potentials from
self-consistent mean-field models}

\author[lanl,lsu]{Z. A. Dupr\'{e}}
\author[lanl,fias]{T. J. B\"{u}rvenich}
\address[lanl]{Theoretical Division, Los Alamos National Laboratory, Los Alamos, NM 87545, USA}
\address[lsu]{Department of Physics and Astronomy, Louisiana State University, Baton Rouge, LS 70803-4001, USA}
\address[fias]{Frankfurt Institute for Advanced Studies, Johann Wolfgang Goethe University, Max-von-Laue-Str. 1,
60438 Frankfurt am Main, Germany}
\begin{abstract}
We present a microscopic model for the calculation of $\alpha$-decay half lives employing potentials obtained from 
relativistic and non-relativistic self-consistent mean-field models. The nuclear and Coulomb potentials are used to obtain
the tunneling probability and, in one model variant, also the knocking frequency. The model contains only one parameter.
We compare this approach employing several modern mean-field para\-me\-tri\-za\-tions to experimental data and to the semi-empirical Viola-Seaborg systematics.
We extrapolate our model to superheavy nuclei where assumptions entering semi-empirical
approaches might lose validity.
\end{abstract}
\end{frontmatter}
\section{Introduction}
Alpha decay was experimentally studied by Rutherford just before the beginning of the 20th century and constitutes
a beautiful example of quantum mechanical tunneling through a (classically impenetrable) barrier.
The $\alpha$-particle or helium nucleus, which is formed within the mother nucleus, tunnels through the Coulomb barrier
created by the protons.
In 1928, after the advent of quantum mechanics, Gamow proposed a semi-classical interpretation for $\alpha$-decay \cite{Gamow}. His model implied that the $\alpha$-decay constant $\lambda$ can be written as
a product of three factors:  the probability $f$ that an alpha particle is formed inside the nucleus ({\em pre-formation factor}), 
the frequency $\nu$ at which the alpha particle knocks against the nucleus's potential well ({\em knocking frequency}), 
and the probability $P$ with which the particle is able to tunnel through the barrier ({\em Gamov factor}), leaving behind the ($Z-2, N-2$) daughter nucleus.
Thus, in this model the $\alpha$ decay constant is given by
\begin{equation}
\lambda = f \times \nu \times P
\label{lambda}
\end{equation}
We note that this approach assumes the statistical independence of these three processes.

The probability of transmission through the barrier, $P=e^\tau$, can be determined by calculating semi-classically in the WKB approximation \cite{VS}
\begin{equation}   
\tau = -2 \int_{r_1}^{r_2} \sqrt\frac{2\mu[V(r)-E]}{\hbar^2}\,dr
\label{transmission}
\end{equation}
The knocking frequency $\nu$ is given by \cite{timedepalpha}
\begin{equation}  
\nu^{-1} = 2 \int_0^{r_1} \sqrt\frac{\mu}{2[E-V(r)]}\,dr,
\label{knocking}
\end{equation}
which, for a constant nuclear potential $V$, reduces to $v/2R$, i.e., the number of times a particle with velocity $v$ (determined from its energy $E$) goes back and forth in a confined space of size $2R$.
In these formulas, $\mu$ is the reduced mass of the daughter-alpha particle system, and $r_1$ and $r_2$ are the points where the energy of the alpha particle is equal to the potential, $r_1$ is the inner, $r_2$ is the outer turning point,
respectively.

Eq. (\ref{knocking}) takes into account the varying contribution of the kinetic to the
total $\alpha$-particle energy which depends on the potential depth at position $r$.
Time-dependent quantum-mechanical calculations \cite{timedepalpha}, however, indicate that the oscillations
of the $\alpha$-particle wave-function within the nucleus have an extremely small amplitude, which might
lead to a better estimate of the number of barrier assaults per second.

The $\alpha$-decay half live $\tau_{1/2}$ is then finally obtained by
\begin{equation}  \tau_{1/2}=\frac{\ln2}{\lambda} 
\end{equation}

A semi-empirical formula for the calculation of $\alpha$ half lives was presented by V. E. Viola and G. T. Seaborg (VS) in 1966 \cite{VS}.  It is based on the Gamow picture and assumes a nuclear potential of uniform depth that 
rises abruptly at the radius of the nucleus, going from this uniform potential to a Coulomb potential, resulting in a cusp at the radius of the nucleus.  Also, the nucleus and therefore the potential are assumed to be spherically symmetric.

The resulting formula reads
\begin{equation}   \log\tau_{1/2}=\frac{aZ+b}{\sqrt{Q_\alpha}} + (cZ+d),
\label{vs}
\end{equation}
yielding the $\alpha$-decay half life $\tau_{1/2}$ in units of seconds.
The experimental input are the charge number $Z$ and the $Q_\alpha$ value (in units of MeV) for a given nucleus $(Z,N)$.
It contains four adjustable parameters $a, b, c, d$. These parameters have been adjusted to the experimental
data on half lives known at that time.
In more recent work, the parameters have been readjusted, taking into account new experimental data on superheavy
elements \cite{IVS}. Table \ref{vs-table} summarizes the values of these parameters and also contains the parameters
of our adjustment made in this paper, see Subsection \ref{subsec_adjustment} for details. The VS systematics
has been used in Ref. \cite{defshe} to calculate $\alpha$ half lives for unknown superheavy nuclei.
We note that there exists a variety of very accurate semi-empirical half live formulas, see for example Refs. \cite{Brown92,Pom83}.
\begin{table}[htb]
\begin{tabular}{r|c|c|c|c}
  & $a$ & $b$ & $c$ & $d$   \\\hline
  original VS     & 2.11329 & -48.9879 & -0.390040 & -16.9543 \\
  readjusted VS  &  1.66175 & -8.5166 & -0.20228 & -33.9069 \\
  our adjustment  & 1.58134 & -2.27889 & -0.23543 & -30.1503
\end{tabular}
\label{vs-table}
\caption{Parameters of the original VS systematics \protect\cite{VS}, its readjustment including data on
superheavy nuclei \protect\cite{IVS}, as well as our adjustment to experimental data from Ref. 
\protect\cite{exp}.}
\end{table}

In this paper, we present a model for the calculation of $\alpha$-decay half lives based
on microscopically calculated potentials from self-consistent mean-field models, namely the relativistic mean-field (RMF) model and the Skyrme-Hartree-Fock (SHF) approach, thus replacing
the simplified assumptions of Eq. (\ref{vs}). In Section 2 we present our approach.
Section 3 describes the adjustment and analysis of our method. We compare its predictions with
both experimental data as well as a new adjustment of the VS systematics. 
In Section 4 we extrapolate to superheavy nuclei and compare the two approaches.
We conclude in Section 5.

\section{Theoretical Framework}
In our approach, which we henceforth denote as $\alpha$-mf, the $\alpha$-decay constant is given by Eq. (\ref{lambda}). The $\alpha$-nucleus potential is given by 
\begin{equation}
V_{\alpha} = 2 \times V_p + 2 \times V_n
\end{equation}
The proton potential $V_p$ and the neutron potential $V_n$ are taken from spherical  self-consistent mean-field
calculations (see below). Thus, $V_\alpha$ contains the sum
of the nuclear and the Coulomb potential. 
In the relativistic models, the $V_\alpha$ potential is obtained
as the sum of the scalar and vector potentials. The $\alpha$-particle is treated as  a spin-saturated
system of two protons and two neutrons, thus it does not feel the spin-orbit force
that is generated by the strong scalar and vector fields adding up with the same sign in the relativistic approaches,  and put in by hand as an additional term in the SHF approach.

The transmission probability and the knocking frequency, given by
Eqs. (\ref{transmission}) and ({\ref{knocking}), respectively, are calculated by numerical
integration using the mean-field potentials. While the bosonic and structure-less $\alpha$-particle picture can be expected to work
for the tunneling process as long as the Compton wavelength of the $\alpha$-particle is
small compared to the width of the barrier, the 'knocking' process of the $\alpha$-particle within
the Coulomb barrier is certainly more complicated  due to the fermionic nature of the nucleons
and the necessary anti-symmetrization between the wave functions of $\alpha$-particle and nucleus.
Still, we may expect that these corrections will not affect too much the overall performance of our model. 

Our approach contains only one free parameter which we denote by $\bar{f}$ and which is related to $f$ in
Eq. (\ref{lambda}). It needs to be fitted to experimental data.
This parameter not only in an average way
describes the formation probability, but also absorbs all errors and approximations made in the calculations of
$P$ and $\nu$. Interpreting it as an (average) probability, it should be less than or equal to $1$. According to Ref. \cite{SKRB} a pre-formation factor should be below $0.1$.
In order to estimate the accuracy of the knocking frequency calculation, we also
build {\em combined parametrizations} with the combined parameter $c = \bar{f} \times \nu$,
in which only the transmission probability is computed numerically.
 
We note that with the knowledge of the
single-particle wave functions, it is possible to calculate the pre-formation factor, see
for example Ref. \cite{preform1}  for a calculation using Nilsson model single particle wave functions
and Ref. \cite{cluster} for a parameter-free approach for calculating the $\alpha$-decay half live of 
$^{212}$Po employing a combined shell and cluster model.
Complementing our model by such a pre-formation calculation would lead to 
a fully microscopic, parameter-free approach, which is, however, beyond the
scope of this paper and will be dealt with in forthcoming work.
The assumption of spherical potentials in our approach loses its validity for deformed nuclei.
It is, however, also used in the Viola-Seaborg systematics and appears to work
well for life-time predictions. Future work will include
potentials taken from deformed mean-field calculations \cite{Dup2}.

The mean-field $\alpha$-potential corrects the simplified potential of other approaches by providing a smooth surface and $r$-dependent interior nuclear potential, and eliminates
the spike at the transition from the nuclear to the Coulomb part, see Fig. \ref{potentials} (the authors of Ref. \cite{Buc92} use a non-self-consistent potential with a smooth surface). Also, its radius automatically possesses the
correct mass-dependence. The accuracy of our model tests the asymptotics of the Coulomb contribution to the mean-field potentials stemming from the relativistic mean-field calculations up to typical radii of $r \approx 70$~fm. The mean-field potentials -- due to their self-consistent interdependence of proton and neutron densities -- automatically introduce a dependence on 
neutron number, i.e., an isospin dependence. For each nucleus, the input to our model is the $V_\alpha$ potential from the daughter nucleus (assuming that the $\alpha$ particle has already formed), the $Q_\alpha$ value and the
mass of the nucleus. The potential is being represented on a radial 1-D grid in coordinate space. The classical turning points
$r_1$ and $r_2$ are determined and then used as boundaries for the numerical computation of transmission and knocking frequency according to Eqs. (\ref{transmission}) and (\ref{knocking}). Because of the uncertainties related to the calculation
of odd-even and odd-odd nuclei, we consider only even-even nuclei in this study.

The self-consistent mean-field models used in this work are on the one hand two variants of the relativistic mean-field model, one employing contact interactions (RMF-PC)
, and the other variant employing finite-range boson fields (RMF-FR),
and on the other hand the Skyrme-Hartree-Fock (SHF) approach. We refer the reader to Refs.  \cite{Review1,PG89,Advances,NHM}
for a detailed discussion of these models. 
In modern terms, self-consistent mean-field models are approximations to the exact
many-body functional in the spirit of density functional theory \cite{Drei90}, their interaction
terms governed by principles of effective field theory such as naive dimensional analysis.
Thus, these models are able to absorb various ground-state correlations and to incorporate physics beyond the (literal) mean-field approximation \cite{furn1}.

The models used here contain
between 6-12 parameters plus two pairing strengths (for protons and neutrons separately) for BCS pairing with a density-independent $\delta$-force.
The parameter sets employed in this work are NL3 \cite{nl3} and NL-Z2 \cite{nl-z2} for RMF-FR, PC-F1 \cite{BOSS} for RMF-PC, and SkI3 \cite{ski3} and SLy6 \cite{sly6} for SHF, which are modern mean-field forces that deliver great predictive power.
They result from careful adjustments to nuclear ground-state observables, e.g., binding energies, rms charge and diffraction radii, and surface thicknesses \cite{BOSS}. 
The nucleon single-particle wave-functions are calculated
self-consistently on a grid in coordinate space employing matrix multiplications in Fourier
space for the derivatives. To obtain the solution with minimal energy, the damped gradient-step
method is used \cite{gradient}. Calculations with both models share the same
basic numerical routines, thus numerical differences do not interfere
with differences in physics for model comparisons.  The center of mass correction is used as in the adjustment
procedure of the mean-field parametrization.

Since the $\alpha$-particle has a finite extension, in addition we test another model variant by
folding its parametrized density distribution with the mean-field alpha potential, i.e.
\begin{equation}
V_{\alpha}^{fold.}(\vec{R}) = \int \rho_\alpha(\vec{r}-\vec{R}) \, V_\alpha(\vec{r}) \, d^3 r 
\end{equation}
The radial density parametrization of the $\alpha$-particle is taken from Ref. \cite{alpha-density} and is given by
\begin{equation}
\rho_\alpha (r) = 0.4229 \times e^{- 0.7024 r^2},
\end{equation}
where $\rho_\alpha (r)$ is given in units of fm$^{-3}$, and the radius $r$ needs to be specified in units of fm.
The effect of the folding procedure on the mean-field potential is shown in Figure \ref{potentials}.

Our $\alpha$-decay model contains no assumptions on the nuclear potential, except for the calculation of the potentials in
spherical symmetry. Thus, provided that the
potentials from the self-consistent mean-field models are realistic, it should lead to more reliable
extrapolations to exotic and superheavy nuclei compared to semi-empirical models. 
We note that these potentials automatically possess -- due to the adjustment of the mean-field models to density-related ground-state observables  -- the correct mass dependence of the potential radius
and also information on the surface region of the nucleus. These need to be parametrized in
more schematic approaches.

We would like to point out possible extensions of the $\alpha$-mf approach, most of
them going clearly beyond the scope of semi-empirical approaches. The microscopic
calculation of the pre-formation factor has already been mentioned. Another enhancement would be the usage of 2-dimensional potentials from axially deformed nuclei. In the case of shape coexistence of nuclei, the dependence of the decay width on
the prolate or oblate minimum can be investigated. Furthermore, decay widths stemming from the ground state or an isomeric state (shape isomer) of a heavy nucleus can be computed.
The description of fine structure
in $\alpha$-decay will be an interesting application, see for example Ref. \cite{fine}.
The model could be enhanced still further by replacing the WKB approximation
with numerical calculations of the stationary or time-dependent Schr{\"o}dinger equation.

\begin{figure}[htb]
\centerline{\epsfxsize=10cm \epsfbox{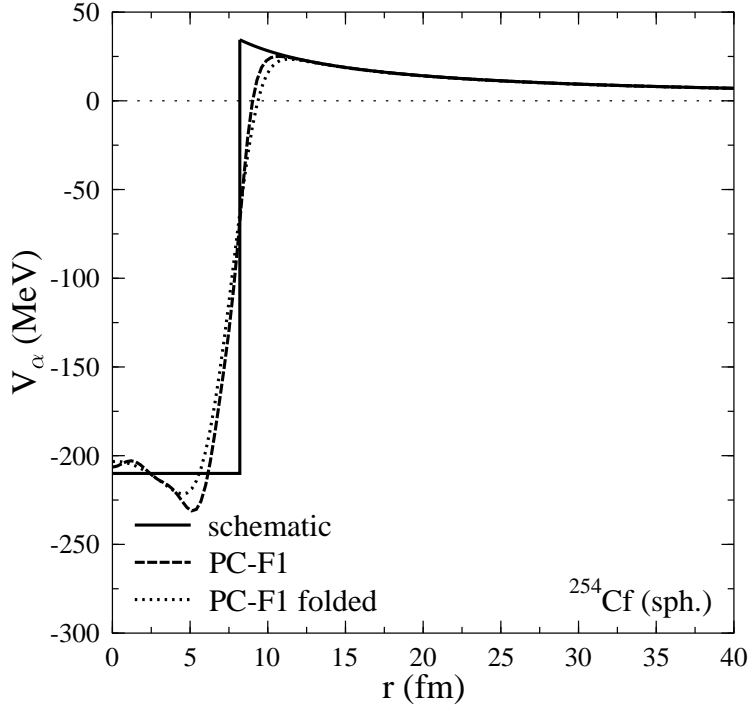}}
\caption{Alpha-nucleus potential obtained from a mean-field calculation with PC-F1 (dashed line),
the same potential but folded with the $\alpha$-particle density (dotted line), and a schematic
potential according to the VS systematics (full line) for the nucleus $^{254}$Cf
}
\label{potentials}
\end{figure}
\section{Results}
\subsection{Adjustment}
\label{subsec_adjustment}
For the model adjustment and comparison, we use recent $\alpha$-decay data from Ref. \cite{exp}.
Since the experimental data varies over thirty orders of magnitude (from $1.1 \times 10^{-7}$~s to $2.2 \times 10^{23}$~s) we have performed a logarithmic  $\chi^2$ fit to adjust the model parameter.  The logarithmic error of a given data point is given by
\begin{equation}   \chi = \log \Big( \frac{\tau_{1/2, th}}{\tau_{1/2, exp}} \Big)
\end{equation}
Computing errors in this way measures how many orders of magnitude the calculated point is away from the experimental value (for example, $\chi = 2.5$ means that the calculated value is 2.5 orders greater than the experimental value, while $\chi = -3$ means that the calculated value is 3 orders of magnitude less than the experimental value). 

To obtain a fair comparison between $\alpha$-mf and VS systematics, we have readjusted the VS systematics to exactly the same data
that has been used in the adjustment of our model \cite{exp}, employing the Monte-Carlo methods
described in Ref. \cite{BOSS} (down-hill methods got stuck very easily in adjacent local minima). 
The Monte-Carlo optimization was finally followed by a down-hill minimization, driving the
parameter vector to the bottom of the (local) minimum. The resulting parameters are shown in Table \ref{vs-table}.
Note that this adjustment has been started with the parameters of the readjusted VS formula.
The usage of Monte-Carlo techniques has reduced $\chi^2$ considerably. However, there might
very well exist several local minima, of which some could correspond to models with even better
predictive power. 

The $\chi^2$ values of the $\alpha$-mf approach employing different mean-field forces are displayed in table \ref{var_forces}. In the following we implicitly refer to $\chi^2$ per point when we write $\chi^2$. 
The $\chi^2$ value of our VS systematics adjustment is included as well.
We see that the Skyrme forces SkI3 and SLy6 in three of four cases deliver a $\chi^2$ value that is below $\chi^2$ of
this adjustment of the VS systematics, while the relativistic models have slightly larger
values. This model difference is quite interesting since it yields information
on the (otherwise unobservable) potentials of the different approaches. The SHF potentials
(for both Skyrme forces) lead to better predictive power compared to RMF. 
This model difference, that might be related to the effective masses, radial dependences, and Coulomb exchange (absent in RMF) deserves further attention in future investigations.
The model variants with the combined parameter $c$ (absorbing both preformation factor and knocking frequency into one parameter) perform better in most cases. The differences are
not dramatic, however, indicating that the explicit calculation of $\nu$ is consistent with
our model assumptions. The (average) preformation factors $\bar{f}$ are smaller than one and thus can be interpreted as probabilities. They also agree with the 
estimates from Ref. \cite{SKRB}. 

The model variant employing the folded potentials from calculations with PC-F1 performs slightly better when explicitely computing the knocking frequency, and slightly worse when absorbing it into the
parameter $c$. This might be connected to the fact that in the folding case we still use the simple reduced mass
in the calculation of the tunneling probability, while the mass parameter is certainly different for
an extended object. The parameter $\bar{f}$ for the folded PC-F1 potential is a factor of 10 smaller than the value without folding,
which is related to the increase of the tunneling probability in the folded case, see Fig. \ref{potentials}. 

We have determined the $\chi^2$ values of the previous fits of the VS systematics using the experimental data from Ref.\cite{exp}.
The original Viola-Seaborg method yields $\chi^2=2.57$, and the improved Viola-Seaborg method $\chi^2=1.27$.

\begin{table}[htb]
\begin{tabular}{l|c|c|| c| c}
force & $\bar{f}$  &  $\chi^2$ &   c (1/s) & $\chi^2$ \\\hline
NL3  & 0.028   & 0.21     &  9.78e19    &  0.21 \\
NL-Z2 & 0.014  & 0.17   & 4.57e19    & 0.16 \\
PC-F1 &   0.080 & 0.24       &     3.09e20        & 0.22                         \\
PC-F1, folded & 0.006 &0.23  & 1.95e19     & 0.24 \\
SkI3  &  0.087    & 0.17  & 3.46e20    & 0.15 \\
SLy6  & 0.075    &  0.15   & 2.85e20    & 0.14 \\ \hline
our VS fit   &   &   0.16 &  & 
\end{tabular}
\caption{The parameter $\bar{f}$ and corresponding $\chi^2$ values for the model variant with explicit calculation of the knocking frequency (columns 1 and 2), and the combined parameter $c$ and corresponding $\chi^2$ values for the combined variant (columns 3 and 4), for the mean-field forces as indicated. For comparison, the $\chi^2$ value of our VS systematics adjustment is shown.}
\label{var_forces}
\end{table}
\subsection{Analysis}
In Figure \ref{megapot} we compare half lives between $\alpha$-mf employing the relativistic mean-field parametrization PC-F1 and experimental data (since the different mean-field approaches
have similar predictive power, comparisons of other forces with experiment yield similar figures).
Note that the $\alpha$-mf approach delivers a very good overall agreement
with $|\chi| < 0.3-0.6$ in most of the cases, i.e., the deviations are well below an order of
magnitude. These errors have predominantly positive signs
corresponding to an over-estimation of the $\alpha$ life-times, i.e., an over-estimation of stability
of these nuclei with respect to $\alpha$-decay. Negative errors are
visible at and near shell closures, see the discussion below. The accuracy of our approach
does not appear to depend strongly on the mass number of the decaying nucleus, indicating that
the mass number dependence is correctly reproduced in the mean-field potentials. The best agreement,
however, is achieved for the actinides.

\begin{figure}[htb]
\centerline{\epsfxsize=14cm \epsfbox{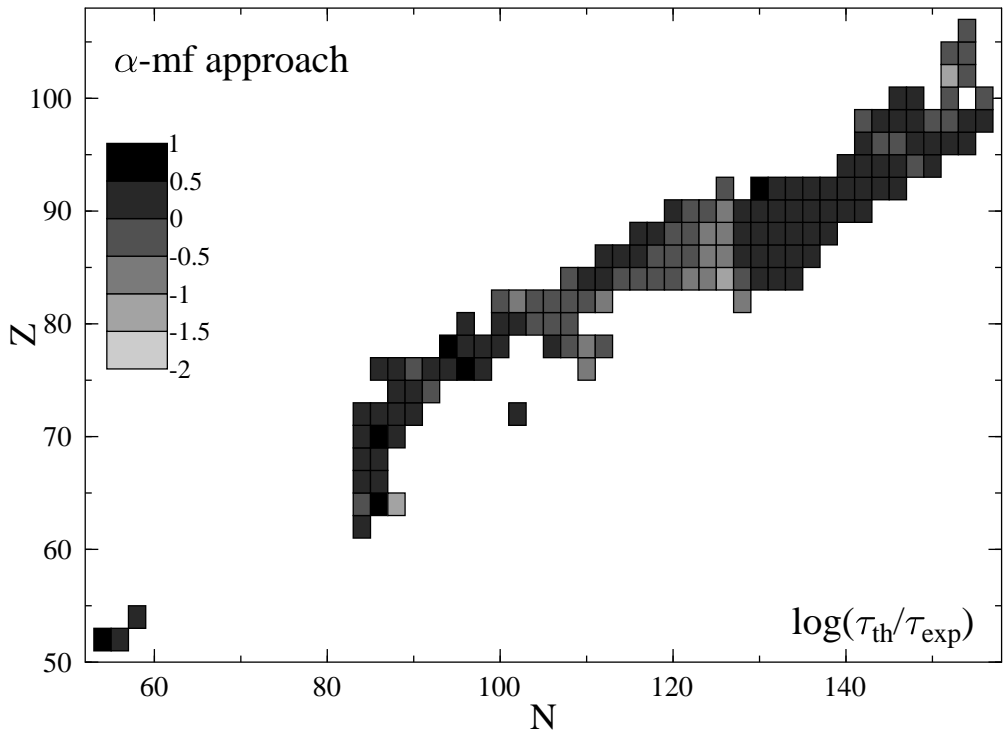}}
\caption{Logarithmic error of our approach with PC-F1 as a contour plot versus neutron number (x-axis) and proton number
(y-axis). Each square corresponds to an even-even nucleus.}
\label{megapot}
\end{figure}
\begin{figure}[htb]
\centerline{\epsfxsize=14cm \epsfbox{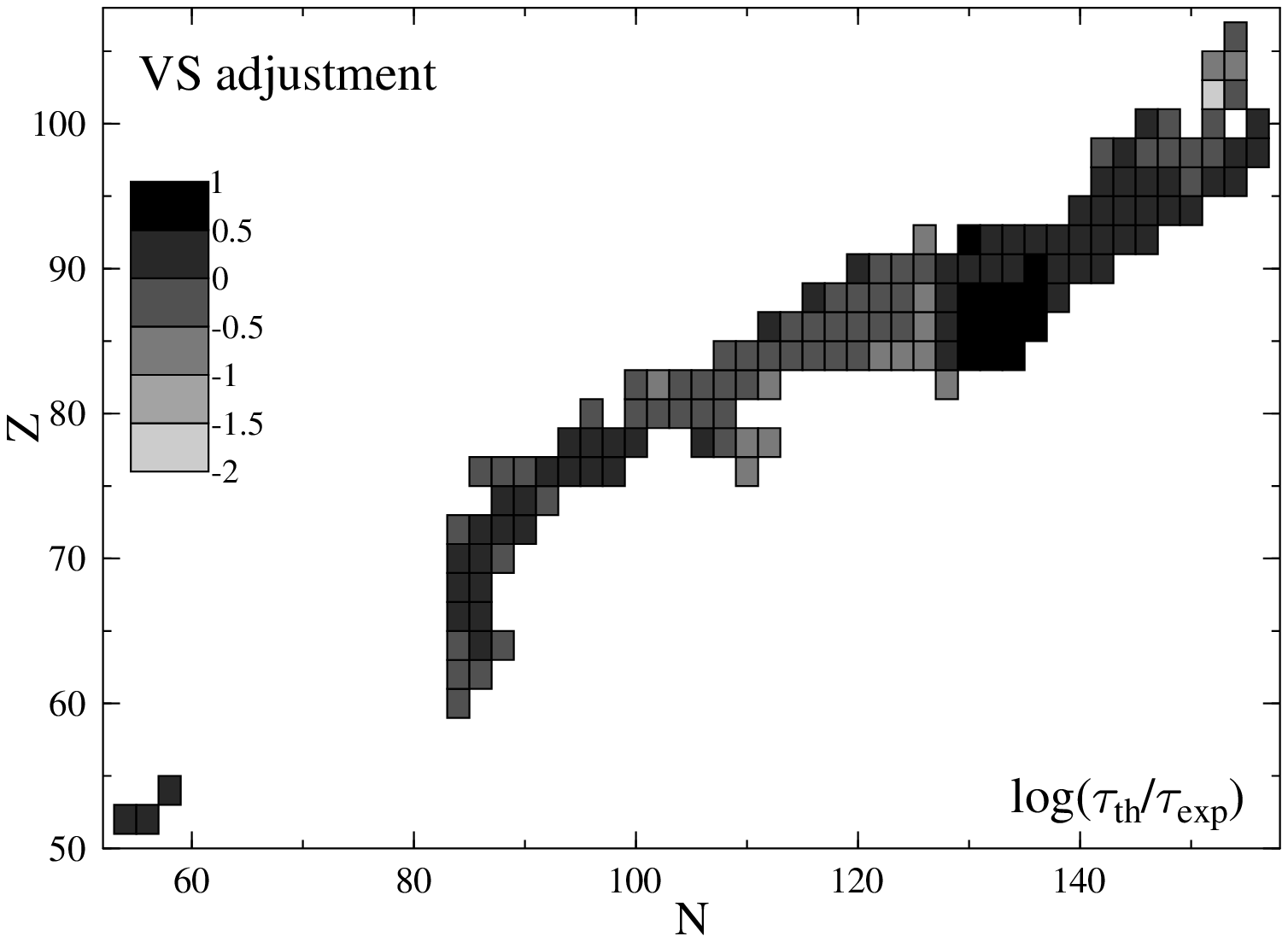}}
\caption{Logarithmic error of the newly adjusted Viola-Seaborg systematics as a contour plot versus neutron number (x-axis) and proton number
(y-axis). Each square corresponds to an even-even nucleus.}
\label{vs_fit}
\end{figure}

We would like to examine our predictions
for indications of physics not (yet) accounted for in our model. We can expect to see deviations
originating from i) shell effects and ii) deformation effects. Let us first examine shell effects.
At the magic neutron number $N=126$, for $Z=84-90$ isotopes we notify a jump in the logarithmic error:
$\alpha$-mf overestimates the stability of $N=128$ isotones, while the stability of $N=126$
isotones is underestimated (indicated by a negative error).  We see a similar effect at the magic
proton number $Z=82$: $\alpha$-mf underestimates stability. This trend can be nicely observed for the $N=128$ isotones. As the proton number drops off towards 82, the nuclides become more and more stable compared to our predictions.
Even though shell effects are taken into account to some extent via the experimental
$Q_\alpha$ values, the computation of the tunneling probability cannot reproduce them sufficiently.
They would enter, however, in a microscopic calculation of the preformation amplitude.

Another cause of trends is the deformation of nuclei.  Our current model -- as does the VS systematics -- assumes spherically symmetric nuclei. It is interesting that the assumption of spherical potentials works so well both in semi-empirical approaches and in our model.
However, most nuclei for which we predict $\alpha$ half lives 
are deformed. Thus, for axially deformed nuclei the Coulomb barrier is 2-dimensional, leading to a more
complicated tunneling dynamics (see Refs. \cite{tal1,tal2} for a time-dependent description of 2-dimensional tunneling in proton emission).
For the most part, our predictions do not display a random distribution pattern, but rather show clusters where the model is more accurate and clusters were it is less accurate.  For example, nuclei with atomic numbers close to $Z=88$ and $N=136$ are less stable than our model predicts, and around $Z=102$ and $N=152$ there is a region of extra stability that our model does not yet account for.  It is not possible to tell directly from our data whether the deformation effect results in making nuclei appear to have extra stability or extra instability, due to the masking effect of optimizing the (average) pre-formation factor. We believe, however, that accounting for deformations in nuclei shall improve our model.

Comparing the predictions of the $\alpha$-mf approach (Fig. \ref{megapot})
with the ones from our VS systematics adjustment (Fig. \ref{vs_fit}), we detect
a few differences. Firstly, nuclei with larger errors with respect to the surrounding
ones are distributed differently in both approaches. Secondly, there are small regions 
of nuclei which are described better in each model, for example the nuclei centered
around $Z=86, N=130$ have larger errors within the VS systematics, while nuclei at or next to the $\alpha$-decay
chain starting at $Z=90, N=120$ are more accurately described by the VS model. Moreover,
we detect similar wrong trends that can be attributed to missing deformation and/or shell structure
features in the VS systematics.

In the next Section, we extrapolate with $\alpha$-mf employing SLy6 to superheavy nuclei
and compare with the predictions of the VS systematics.

\section{Extrapolation to superheavy elements}
\begin{figure}[htb]
\centerline{\epsfxsize=14cm \epsfbox{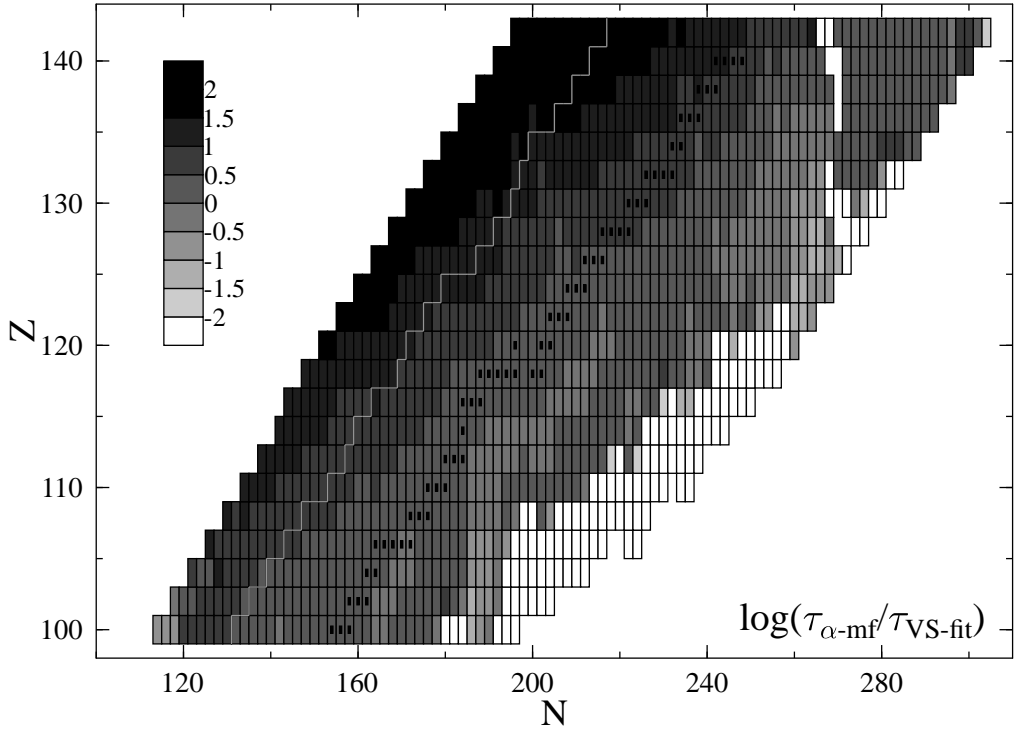}}
\caption{Logarithmic ratio of the predicted life time from $\alpha$-mf over the predicted life time of
our Viola-Seaborg adjustment systematics as a contour plot versus neutron number (x-axis) and proton number
(y-axis). Each square corresponds to an even-even nucleus, nuclei without a converged solution have been omitted. The black squares indicate the valley
of $\beta$ stability, the grey line corresponds to the two-proton dripline.}
\label{she}
\end{figure}

We employ the $\alpha$-mf approach with the force SLy6, which has performed best in
terms of $\chi^2$, for an extrapolation to superheavy nuclei. We compare its predictions
to the predictions from our adjustment of the VS systematics. The $Q_{\alpha}$ values are
obtained from axially deformed ground-state calculations with SLy6, similarly
to Refs. \cite{Typel1,defshe}. The 
potentials are obtained from calculations in spherical symmetry. For both models,
the $Q_{\alpha}$ values calculated with SLy6 are used, since a) experimental data in that region
is very scarce (especially for even-even nuclei) and b) we aim at a pure model comparison. 

The logarithmic ratio of the two approaches can be seen in
Fig. \ref{she}. The difference between the two approaches remains quite small
(below 0.1) for nuclei close to $\beta$-stability. Also, both models
show no visible difference in the error with mass number, a feature we have
seen before for the known elements.
However, an isovector trend appears in the figure between the two methods.
With decreasing $N-Z$ ratio, the $\alpha$-mf approach predicts increasingly
larger life times. Since both approaches use the same -- calculated --
$Q_\alpha$ values, the isospin dependence has its origin in the
potential shapes used in $\alpha$-mf which introduces a dependence on the
neutron number. This isovector trend is absent in the VS systematics.

For nuclei with masses beyond superheavy nuclei $(A \ge 292)$, various groups have theoretically found 
semi-bubble and bubble structures \cite{diet76,diet98,rutz97,dech99}, i.e., the density in the interior of the nucleus is reduced or
even completely suppressed, leading to nuclear systems with an inner surface. Because of this structure, these systems possess potentials
that differ to a great extent from usual nuclear potentials. Thus, they will constitute an interesting application
for $\alpha$-decay models based on self-consistent potentials \cite{Dup2}.

\section{Conclusions}

We have presented a one-parameter model for $\alpha$-decay that can be used to predict life times of $\alpha$-emitters.  This model can be applied successfully over a large range of nuclei. Its accuracy is comparable to or even slightly better than semi-empirical
approaches. This has been demonstrated
by a new adjustment of the Viola-Seaborg formula.
However, universality and extensibility are the decisive features of this model. It can be applied to exotic and superheavy nuclei, in which the assumptions of (very successful) semi-empirical formulas about the shape of the potential lose more and more validity, given that the potentials
from the self-consistent mean-field models are still realistic. 
By employing a variety of recent mean-field parametrizations, we have found that $\alpha$-nucleus potentials from the Skyrme-Hartree-Fock method yield slightly better results compared
to the RMF model potentials. 

An extrapolation to superheavy elements has shown that the $\alpha$-mf approach displays
an isospin dependence of life times which is absent in the Viola-Seaborg approach. This point deserves
further attention as new data become available.

The model is simple in terms of required parameters.  While the Viola-Seaborg systematics contains four parameters that need to be simultaneously optimized, it contains only one adjustable parameter, the (average) pre-formation (or a combined) factor.  Furthermore, our model has ample room for improvement. One potential improvement would be to  microscopically calculate the pre-formation factors using nuclear single-particle and $\alpha$-particle wave-functions, yielding a fully microscopic, parameter-free approach. Other enhancements and applications involve multi-dimensional potentials and fine-structure $\alpha$-decay calculations.
\section*{Acknowledgements}
Z. A. D. and T. J. B. would like to thank N. Magee, L. Collins, D. James and S. Seidel for the organizing of the
Los Alamos Summer School 2004.
T. J. B. would also like to thank D. G. Madland, C. M\"uller, P.--G. Reinhard, O. Serot, and P. Talou for helpful comments.
This work was supported by the U.S. Department of Energy.

\end{document}